 \documentclass[preprint2]{aastex}

\newcommand{\etal}{{\it et al.}\ }
\slugcomment{Revised version - Nov '01}

\shorttitle{UV Properties of Primveval Galaxies}
\shortauthors{A. Buzzoni}

\begin{document}

\title{Ultraviolet properties of primeval galaxies:\\
       theoretical models from stellar population synthesis}

\author{Alberto Buzzoni}
\affil{Telescopio Nazionale Galileo, A.P. 565 38700 S/Cruz de La Palma (Tf), Spain, \and
Osservatorio Astronomico di Brera, Milano, Italy}

\email{buzzoni@tng.iac.es}

\begin{abstract}
The ultraviolet luminosity evolution of star-forming galaxies is explored from
the theoretical point of view, especially focusing on the theory of UV energetics
in simple and composite stellar populations and its relationship to the
star formation rate and other main evolutionary parameters.

Galaxy emission below $\lambda < 3000$~\AA\ directly correlates with actual
star formation, not depending on the total mass of the system. A straightforward calibration
is obtained, in this sense, from the theoretical models at 1600, 2000 and 2800 \AA, and a full
comparison is carried out with IUE data and other balloon-borne observations for local galaxies.

The claimed role of late-type systems as prevailing contributors to the cosmic
UV background is reinforced by our results; at 2000 \AA\ Im irregulars are found in fact
nearly four orders of magnitude brighter than ellipticals, per unit luminous mass.

The role of dust absorption in the observation of high-redshift galaxies is assessed,
comparing model output and observed spectral energy distribution of local galaxy samples.
Similar to what we observe in our own galaxy, a quick evolution in the dust environment might
be envisaged in primeval galaxies, with an increasing fraction of luminous matter that would escape the 
regions of harder and ``clumpy'' dust absorption on a timescale of some $10^7$~yr, comparable with the 
lifetime of stars of 5--10~M$_\odot$.

\end{abstract}


\hfill {\it In ricordo di mio padre, per}

\hfill {\it tutto il bene che mi ha dato.}
                               
\section{Introduction}
The recent major improvement in the observation of the deep Universe, both with HST and ground-based 
telescopes of the new generation, has greatly increased our chance to probe the cosmological model 
by venturing to search for primeval galaxies. 
When observing in the optical or infrared range, however, we are probing restframe ultraviolet
(UV) emission of high-redshift objects, and we should therefore rely on confident recognition criteria 
in this wavelength range in order to pick up distant galaxies and track their evolution back in time.

Because of atmosphere opacity, nearby galaxies are still poorly explored in the extreme UV range, and their data
cannot fully match high-$z$ observations (Lanzetta \etal 1996; 
Massarotti \etal 2001a). In any case, such an empirical approach would neglect evolution, while sample 
selection and Malmquist bias could give rise to a misleading interpretation of the high-redshift data 
(Buzzoni 1998, 2001; Adelberger and Steidel 2000). 
For all these reasons, galaxy models may prove to be a more effective tool to match primeval galaxy
evolution, although within the limits of the established theoretical scenario.

The observation of UV emission of star-forming galaxies at cosmological distances, in
the Hubble Deep Field (Williams \etal 1996) or in other deep surveys (Lilly \etal 1996; 
Connolly \etal 1997), has recently led to important advances in the study of the cosmic star formation 
(Madau \etal 1998; Steidel \etal 1999; Massarotti \etal 2001b) fueling the current debate on 
the epoch of galaxy formation. In the light of these results, in this work my aim is to assess in more 
detail the expected UV properties of primeval galaxies. In particular, I focus on the 
theory of UV energetics in stellar populations, in order to explore its dependence on the 
star formation rate (SFR) and other main evolutionary parameters. This will rely
on a previous theoretical framework for evolutionary population synthesis (Buzzoni 1989, 1995, 1998).
To some extent, my analysis is complementary to the work of Leitherer \etal (1999), 
as I track here late evolution of star-forming galaxies, when morphology 
begins to differentiate late- and early-type systems along the Hubble sequence.

My discussion will first consider, in Sec.~2, the UV evolution of a simple stellar population
(SSP). This will provide the basic theoretical tools to derive an absolute calibration of
galaxy UV luminosity and actual SFR, and to study its possible dependence on metallicity.
A possible scheme for the spectral evolution of star-forming galaxies of different morphological type
is outlined in Sec.~3, where I compute a new set of synthesis models and
compare, in Sec.~4, the theoretical output with IUE observations and other UV data available for local galaxies.
The problem of dust absorption is also reviewed in the latter section 
discussing its possible impact on the evolutionary scenario of high-redshift galaxies.
Our relevant conclusions are finally summarized in Sec.~5.

\begin{figure}[t]
\resizebox{\hsize}{!}{\includegraphics{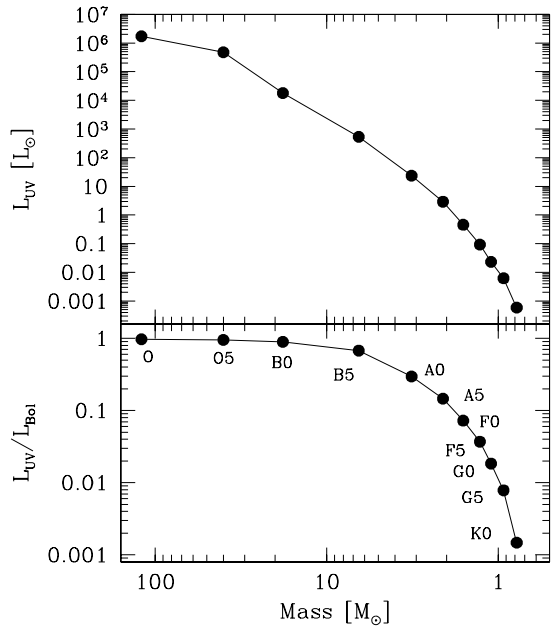}}
\caption{
UV $L$--$M$ relation for MS stars (MK class V). The top panel reports the emission for
$\lambda < 3000$~\AA, in bolometric solar units. The relative fraction of ultraviolet
to bolometric is displayed in the bottom panel. Kurucz' (1992) model
atmospheres have been used to compute synthetic spectra. The calibration for the
120~M$_\odot$ star is from Bressan \etal (1993).
}
\label{stars}
\end{figure}

\section {Fundamentals of UV luminosity in stellar populations}

Contrary to the case of optical and infrared evolution, the ultraviolet luminosity of 
star-forming galaxies is largely dominated by short-lived stars of high mass (Kennicutt 1998).
This makes population synthesis models much simpler, because hot main sequence (MS) stars are 
the prevailing contributors to the galaxy energetic budget at short wavelengths.
In a more refined analysis, however, the input from low-mass stars (M~$\lesssim 1$~M$_\odot$)
should also be conveniently sized up, since these stars smoothly cumulate over the entire galaxy life 
and could reach hot temperatures at some stage of the post-MS evolution.
The SSP theory (Renzini and Buzzoni 1986; Buzzoni 1989) provides an effective tool 
to consistently assess the general UV properties of stellar populations. As we will see,
the relevant results for the SSP case can also easily apply to the study of star-forming galaxies.

An illustrative summary of the integrated luminosity emitted shortward of 3000 \AA\ by MS stars of 
different spectral type is reported in Fig.~\ref{stars}. The plot is based on the Kurucz (1992) model atmospheres,
adopting a standard temperature-luminosity-mass calibration for class V stars 
of solar metallicity as from Allen (1973).\footnote {Allen's (1973) classical compilation for MS stellar parameters 
is also a suitable match to more sophysticated theoretical calibrations, such as, e.g. those
deriving from the Padova (Bressan \etal 1993) and Geneva (Schaller \etal 1992) stellar tracks.}
One sees from the figure that at least half the bolometric luminosity of stars more massive than
$\sim 5$~M$_\odot$ is spent in the ultraviolet. 
For stars earlier than $F0$ (M~$\gtrsim 2$~M$_\odot$), a simple $L$-$M$ calibration can be derived
such as $L_{UV} = K\ M^\alpha$, with $(K, \alpha) = (0.21, 4.0)$ for $\lambda < 3000$~\AA\ 
(expressing both $L$ and $M$ in solar units).
At lower mass, the slope of the relation quickly steepens, and the UV contribution drops to 
nominal values for $K$ stars.

As far as a SSP is considered, with stars distributed in mass according to a power-law IMF
such as $dN = A\,M^{-s}dM$ ($s = 2.35$ for the Salpeter standard case),
the integrated MS luminosity results:
\begin{equation}
L_{MS} = A\,K \int_{M_l}^{M_{TO}}m^{\alpha-s}\ dm.
\label{eq:int_ms}
\end{equation}
where $A$ is a normalization factor that scales with the SSP total mass, while M$_l$ 
is the lower limit for the IMF. The integral is constrained by the mass of the turn-off (TO) stars (M$_{TO}$),
that leave the MS. This actually marks the ``clock'' of the SSP, giving the time dependence of $L_{MS}$.
Recalling that $L_{TO} = K\,M_{TO}^\alpha$, and provided that $s < (1+\alpha)$ and $M_{TO} \gg M_l$, after some 
appropriate substitutions we have:
\begin{equation}
L_{MS} = {A\over{1+\alpha-s}}\ M_{TO}^{1-s}L_{TO}.
\label{eq:lms}
\end{equation}
Equation~(\ref{eq:lms}) simply relates total MS luminosity to the TO
luminosity of the evolving stars in the SSP.

\begin{figure}[t]
\resizebox{\hsize}{!}{\includegraphics{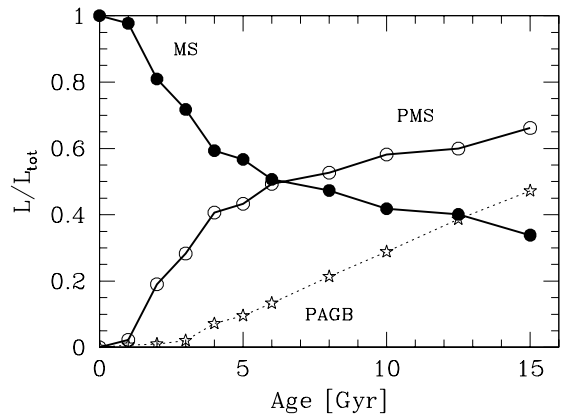}}
\caption{
The SSP luminosity evolution in the UV wavelength range ($\lambda < $ 3000 \AA)
according to Buzzoni's (1989) synthesis code. A solar metallicity and a Salpeter IMF
are assumed.
The relative contribution from MS ($\bullet$) and Post-MS ($\circ$) stars is compared.
The luminosity fraction provided by Post-AGB stars alone ($\star$) is also singled out.}
\label{uvfrac}
\end{figure}

Detailed SSP evolution for $\lambda < 3000$ \AA\ is computed in Fig.~\ref{uvfrac}, where I extend
to younger ages the original model sequence of Buzzoni (1989) for a Salpeter IMF and solar metallicity.
In this figure, I also single out the contribution from post-AGB stars experiencing the
planetary nebula event for $t \gtrsim 1$~Gyr.
As is well known, post-AGB stars are important contributors to short-wavelength luminosity in old SSPs, 
and may actually constrain the ultraviolet properties of present-day elliptical galaxies
(Renzini and Buzzoni 1986; Burstein \etal 1988; Greggio and Renzini 1990; Yi \etal 1999).
The SSP model of Fig.~\ref{uvfrac} assumes a red horizontal branch (HB) morphology; my choice relies
on the fact that HB evolution quickly turns to redder colors with increasing star mass and/or metallicity.
Although stars with temperatures in excess of 10\,000~K are recognized in some old metal-poor Galactic 
globular clusters (Cacciari \etal 1995), a red HB is always to be expected at younger ages 
(i.e.\ for $t \lesssim 12$~Gyr; cf.\ Fig.~4 of Buzzoni 1989).

The resulting SSP luminosity evolution at three reference wavelength values,
namely, 1600, 2000, and 2800 \AA, is reported in Table~\ref{SSP}
for a Salpeter IMF with an upper cutoff mass of 120~M$_\odot$.

\subsection {UV luminosity and SFR}

\begin{figure}[t]
\resizebox{\hsize}{!}{\includegraphics{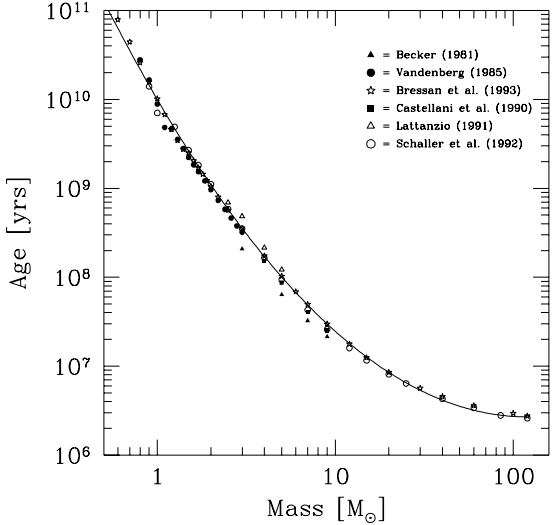}}
\caption{
Theoretical MS lifetime vs.\ stellar mass for $Z_\odot$ according to the evolutionary 
tracks by Becker (1981), Vandenberg (1985), Castellani \etal (1990), Lattanzio (1991), 
Schaller \etal (1992), and Bressan \etal (1993). The solid line is a fit to 
the data according to eq.~(\ref{eq:fit_clock}).}
\label{clock}
\end{figure}

Insofar as MS contribution is the prevailing one in the UV energetic budget,
the SSP luminosity evolution can be approximated by a simple analitical relation linking
eq.~(\ref{eq:lms}) with the theoretical stellar clock.
The MS lifetime for stars in the different range of mass and solar metallicity can be
evaluated from a number of sets of theoretical models. I considered the work of Becker 
(1981), Vandenberg (1985), Castellani \etal (1990), Lattanzio (1991), Schaller \etal (1992), 
and Bressan \etal (1993). 
A useful fit to these data, which spans the whole range of mass, is
\begin{equation}
\log t = 0.825 \log^2(M_{TO}/120) + 6.43
\label{eq:fit_clock}
\end{equation}
(cf. Fig.~\ref{clock}).

By differentiating eq.~(\ref{eq:lms}) and eq.~(\ref{eq:fit_clock}) we obtain
$d\log L_{MS} = (1+\alpha-s)\ d\log M_{TO}$ and $d\log t = 1.65\,\log(M_{TO}/120)\ d\log M_{TO}$,
respectively.
Assuming that $L_{MS} \equiv L_{SSP}$, this leads to
\begin{equation}
L_{SSP} \propto t^{0.6(1+\alpha-s)/\log(M_{TO}/120)} \equiv t^{-\gamma}.
\label{eq:gamma}
\end{equation}
The power index $\gamma$ depends on $M_{TO}$ and $\alpha$, so that it slightly changes with time and wavelength 
(SSP luminosity fades more rapidly at shorter wavelength) but, as a general case, we could verify
from Table~\ref{SSP} that $\gamma \geq 1$ for $\lambda < 3000$~\AA. This feature has important consequences when
linking UV luminosity and galaxy SFR.
In the case of a star-forming galaxy with constant SFR, for example, actual luminosity simply follows as
\begin{equation}
L_{gal} \propto {\rm SFR} \int_{t_{min}}^{t_{gal}} \tau^{-\gamma}d\tau \propto [{\rm SFR}\ t_{min}]\ L_{tmin}^{SSP}
\label{eq:lgal}
\end{equation}
where $t_{gal}$ is the age of the system, and $t_{min}$ is the lifetime of the most massive stars in the IMF.
The previous equation confirms that galaxy UV luminosity eventually does not depend on age, but only on its 
actual SFR (Kennicutt 1998, cf.\ his Fig.~2).

As a consequence, a {\it minimum} UV luminosity is reached by a star-forming galaxy that is the
luminosity of a SSP of total mass $M_{tot} = {\rm SFR} \times t_{min}$, where $t_{min}$ comes from 
eq.~(\ref{eq:fit_clock}) once entering the upper stellar mass of the IMF ($M_{up}$).
A straightforward relation exists between $M_{tot}$ and the SSP scale factor $A$ in eq.~(\ref{eq:lms})
(cf. e.g. eq.~[3] in Buzzoni 1989); making the relevant substitutions, we can write
\begin{equation}
\log L_{min} = \log \phi + \log {\rm SFR},
\label{eq:lmin}
\end{equation}
with
\begin{equation}
\phi = t_{min} {{|2-s|}\over{(1+\alpha-s)}}\Big({{M_{up}}\over{M_{cut}}}\Big)^{2-s}\Big({{L_{up}}\over{M_{up}}}\Big).
\end{equation}
In the previous equation, $M_{cut}= M_l$ if $s>2$ and $M_{cut}=M_{up}$ if $s<2$, while
$L_{up}$ is the TO luminosity of stars of mass $M_{up}$.

\begin{figure}[t]
\resizebox{\hsize}{!}{\includegraphics{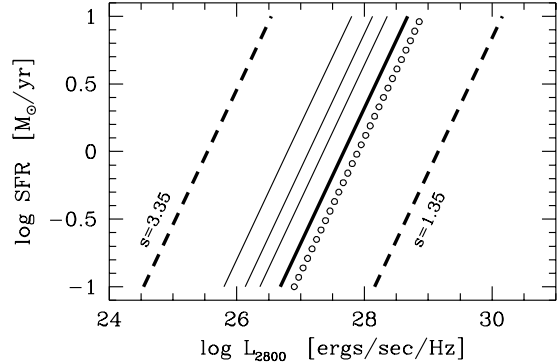}}
\caption{
Theoretical L$_{UV}$ vs.\  SFR calibration at 2800 \AA\ according to eq.~(\ref{eq:lmin}).
Solid lines (from the left to the right) refer to a Salpeter IMF ($s = 2.35$) and $M_{up}
= 40$, 60, 80, and 120~M$_\odot$, with the last case marked in boldface.
The IMF slopes for $s = 3.35$ and 1.35 and $M_{up}$ = 120~M$_\odot$ are shown by the 
two dashed lines.  The Madau \etal (1998) calibration for a Salpeter IMF and 
$M_{up}$ = 125~M$_\odot$ is also reported for comparison (dotted line).}
\label{calib}
\end{figure}

\begin{deluxetable}{lrrr}
\tabletypesize{\footnotesize}
\tablecolumns{4}
\tablewidth{0pc}
\tablecaption{SSP ultraviolet evolution\tablenotemark{(a)}}
\tablehead{
\colhead{Age} & \multicolumn{3}{c}{$\Delta \log L$} \\
\colhead{[Gyr]} & \multicolumn{3}{c}{\hrulefill} \\
   &   \colhead{1600 \AA} & \colhead{2000 \AA} & \colhead{2800 \AA} }
\startdata
\phn 0.003 &  0.00 &  0.00 &  0.00 \\
\phn 0.013 & --2.13 & --1.94 & --1.76 \\
\phn 0.024 & --2.49 & --2.27 & --2.06 \\
\phn 0.046 & --2.85 & --2.60 & --2.35 \\
\phn 0.057 & --2.96 & --2.70 & --2.44 \\
\phn 0.078 & --3.13 & --2.85 & --2.58 \\
\phn 0.10 & --3.27 & --2.96 & --2.68 \\
\phn 0.20 & --3.66 & --3.29 & --2.97 \\
\phn 0.40 & --4.26 & --3.69 & --3.30 \\
\phn 0.60 & --4.94 & --4.00 & --3.52 \\
\phn 0.80 & --5.47 & --4.34 & --3.73 \\
\phn 1.0 & --5.87 & --4.64 & --3.89 \\
\phn 2.0 & --7.04 & --5.51 & --4.37 \\
\phn 4.0 & --7.90 & --6.23 & --4.68 \\
\phn 5.0 & --8.02 & --6.39 & --4.78 \\
\phn 6.0 & --8.03 & --6.56 & --4.88 \\
\phn 8.0 & --8.03 & --6.85 & --5.05 \\
10.0 & --8.00 & --7.11 & --5.19 \\
12.5 & --7.88 & --7.27 & --5.32 \\
15.0 & --7.77 & --7.37 & --5.43 \\
\enddata
\tablenotetext{(a)}{For a Salpeter IMF with upper cutoff mass of 120 $M_\odot$.}
\label{SSP}
\end{deluxetable}

\begin{deluxetable}{ccccc}
\tabletypesize{\footnotesize}
\tablecolumns{5}
\tablewidth{0pc}
\tablecaption{The $\log \phi$ calibration for star-forming galaxies\tablenotemark{(a)}}
\tablehead{
\colhead{Wavelength} & \multicolumn{4}{c}{$M_{up}$} \\
\colhead{[\AA ]} & \multicolumn{4}{c}{\hrulefill} \\
   &   \colhead{120 M$_\odot$} & \colhead{80 M$_\odot$} & \colhead{60 M$_\odot$} & \colhead{40 M$_\odot$}
 }
\startdata
           & \multicolumn{4}{c} {$s=3.35$}\\
     & \multicolumn {4}{c} {\hrulefill} \\
1600 & 25.77 & 25.62 & 25.51 & 25.36 \\
2000 & 25.71 & 25.56 & 25.45 & 25.30 \\
2800 & 25.54 & 25.39 & 25.28 & 25.13 \\
\noalign{\smallskip}
           & \multicolumn{4}{c} {$s=2.35$}\\
     & \multicolumn {4}{c} {\hrulefill} \\
1600 & 27.92 & 27.60 & 27.37 & 27.04 \\
2000 & 27.86 & 27.54 & 27.31 & 26.98 \\
2800 & 27.68 & 27.36 & 27.13 & 26.80 \\
\noalign{\smallskip}
           & \multicolumn{4}{c} {$s=1.35$}\\
     & \multicolumn {4}{c} {\hrulefill} \\
1600 & 29.39 & 28.89 & 28.53 & 28.03 \\
2000 & 29.33 & 28.83 & 28.47 & 27.97 \\
2800 & 29.16 & 28.66 & 28.30 & 27.80 \\
\enddata
\tablenotetext{(a)}{The listed quantity is $\log L_{min}$ in erg~s$^{-1}$Hz$^{-1}$\\
according to eq.~(\ref{eq:lmin}) for a SFR = 1 M$_\odot$yr$^{-1}$.}
\label{zero}
\end{deluxetable}

The expected L$_{UV}$ vs.\ SFR calibration at 2800 \AA\ is shown in
Fig.~\ref{calib}; a fairly good agreement is found with the Madau \etal (1998) results.
As L$_{UV}$ tracks the SFR via the relative number
of high-mass stars, any non-standard IMF could easily be accounted for.
For example, a Scalo (1986) mass distribution would closely resemble in Fig.~\ref{calib}
the case of a power-law IMF with $s \sim 2.5$.
Table~\ref{zero} reports the values of $\log \phi$ at 1600, 2000, and 2800 \AA\ for different IMF slopes 
around the Salpeter value.

\subsection {Metallicity effects}

As metal enrichment is a direct by-product of star 
formation, one would expect primordial galaxies to be essentially metal-poor 
aggregates. For this reason, it is also relevant to assess the validity of our 
L$_{UV}$ vs.\ SFR calibration in case of a non-solar evolutionary scenario
with $Z \ll Z_\odot$.
A change in metallicity would basically affect the MS fuel consumption, modulating
the $\phi$ scaling factor in eq.~(\ref{eq:lmin}) via the term $L_{up}t_{min}$.

\begin{figure}[t]
\resizebox{\hsize}{!}{\includegraphics{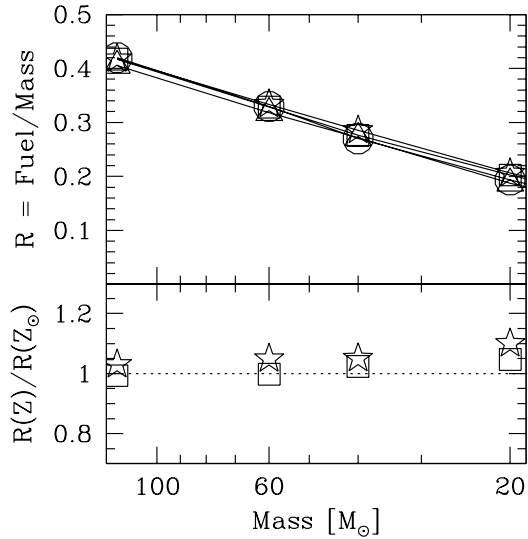}}
\caption{
{\it Top panel:} Fraction of (initial) stellar mass consumed during MS evolution
according to the Padova (Bressan \etal 1993; Fagotto \etal 1994)
and Geneva (Schaller \etal 1992) evolutionary tracks.
The plotted quantity is $R =$~fuel/mass where fuel~$= (L_{ZAMS}\times t_{MS})/(0.007\ c^2)$
converted to Hydrogen-equivalent solar masses.
Padova tracks are for $Z_\odot$ ($\vartriangle$) and $Z = 0.004$ ($\star$)
while Geneva tracks are for $Z_\odot$ ($\circ$) and $Z = 0.001$ ($\sq$).\protect \\
{\it Bottom panel:} Relative change in MS fuel consumption for the Padova 
$Z = 0.004$ tracks with respect to the solar case ($\star$) and, similarly, for the
Geneva $Z = 0.001$ tracks ($\sq$).}
\label{fuel}
\end{figure}

This point is explored in Fig.~\ref{fuel}, in which I compute the equivalent quantity
fuel~=~L$_{ZAMS} \times$~t$_{MS}$ from the Padova (Bressan \etal 1993; 
Fagotto \etal 1994) and Geneva (Schaller \etal 1992) stellar tracks, and study
its absolute and relative variation with changing $Z$. In the top panel of 
the figure, the fraction of initial stellar mass burnt during MS evolution
is reported for the model sequences at solar metallicity and the corresponding metal-poor ones
(namely, $Z = 0.001$ for the Geneva tracks and $Z = 0.004$ for the Padova tracks). 
Compared with the $Z_\odot$ case, metal-poor stars increase 
their MS lifetime but decrease bolometric (and UV) luminosity (cf.\ bottom panel).
The two effects tend to nearly compensate, so that the factor $L_{up}t_{min}$ 
remains almost constant over a wide range of stellar mass.
This secures our L$_{UV}$ vs.\ SFR calibration within just 
a $\pm 10$~\% uncertainty even for quite extreme deviations from the solar metallicity.

\begin{figure}[t]
\resizebox{\hsize}{!}{\includegraphics{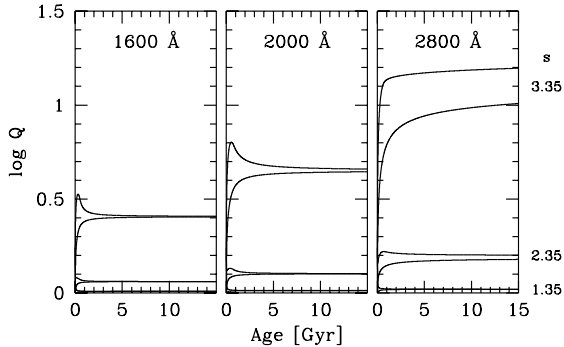}}
\caption{
$Q$ residual luminosity from the bulk of unevolved stars,
after eq.~(\ref{eq:q}) for different IMF power-law slopes
with $s = 1.35, 2.35$, and 3.35, as labeled on the plots.
Each set of curves provides the envelope for a SFR~$\propto t^{-\eta}$
assuming $\eta$ in the range +0.8 (upper envelope) and  --0.8 (lower envelope). An 
upper IMF cutoff mass of 120~M$_\odot$ is adopted.}
\label{q}
\end{figure}

\subsection{Actual and integrated star formation}

Relying on eq.~(\ref{eq:lmin}), a convenient way to parameterize the luminosity
evolution of a star-forming galaxy is
\begin{equation}
\log L_{gal}(t) = \log \phi + \log {\rm SFR}(t) + \log Q(t).
\label{eq:q}
\end{equation}
In addition to the contribution of fresh star formation, the correcting factor $Q(t)$ in the 
equation takes into account the ``extra'' luminosity contributed  by the bulk of unevolved 
low-mass stars that have been accumulating along the galaxy's life. It is interesting to theoretically 
evaluate this term in order to assess its real impact on galaxy spectral energy distribution (SED)
for different star formation histories.

In its general case, the total luminosity of a star-forming galaxy can be computed as a suitable 
convolution of SSP ``building blocks'' with the SFR at the different epochs:
\begin{equation}
L_{gal}(t)  = \int_{t_{min}}^t L_{SSP}(\tau)\ SFR(t-\tau)\ d\tau.
\label{eq:lgal}
\end{equation}
Once $L_{gal}$ is obtained from eq.~(\ref{eq:lgal}), and $L_{min}$ from eq.~(\ref{eq:lmin}),
the $Q$ residual luminosity simply derives from eq.~(\ref{eq:q}).
For the present experiment, it is useful to consider the case of a power-law time decay such 
as SFR~$\propto t^{-\eta}$. The major advantage of this simple parameterization 
is that it can easily account for a wide range of star formation histories in terms of an
age-independent\footnote{At age $t$, SFR~$= C\ t^{-\eta}$ and
its time average is $<{\rm SFR}> = C\ t^{-1}\int_0^t \tau^{-\eta}d\tau$.
Providing that $\eta < 1$, we always have $<{\rm SFR}> = {\rm SFR}/(1-\eta)$, from which
the value of $b$ directly follows, by definition.}
distinctive birthrate
\begin{equation}
b = {\rm SFR}/<{\rm SFR}> = (1-\eta).
\label{eq:birth}
\end{equation}
The value of $\eta$ can be tuned up in order to reproduce galaxy colors 
at present time. For a Salpeter IMF with stars between 
0.1 and 120~M$_\odot$, its allowed range is between $\pm 0.8$ (cf.\ Sec.\ 3). 
With these assumptions, the resulting $Q$ correction is summarized in Fig.~\ref{q}.

After a sharp increase on a few $t_{min}$ timescale, $\log Q$ tends to flatten and proceed steadily 
over the whole galaxy life. As expected, the luminosity correction becomes more important at longer 
wavelengths, and for a dwarf-dominated IMF (i.e.\ $s \gg 2.35$). However, for a Salpeter IMF, and even at 
2800 \AA, its value is less than $\sim 0.2$~dex, confirming that UV luminosity of star-forming galaxies 
is only marginally modulated by the past star formation history. When moving to shorter wavelengths, 
eq.~(\ref{eq:lmin}) always tends to a fair estimate of galaxy total luminosity, and SED 
will closely resemble that of the hottest composing stars (as $L_{min} \propto L_{up}$ in the equation).

\section {Ultraviolet vs.\ optical luminosity}

As I showed in previous section, past star formation history is better traced by the galaxy SED at 
longer wavelengths.
Combining ultraviolet and optical observations, such as, for example, in the $B$ and 
$V$ bands, could therefore set valuable constraints on the stellar birthrate for
galaxies along the whole Hubble morphological sequence (Larson and Tinsley 1978).

\subsection {The galaxy synthesis models}

To further investigate this important issue, this section introduces a simple family of galaxy models
consisting of a spheroid and disk stellar component. Apart from minor details, this is basically the set 
of templates adopted by Massarotti \etal (2001a,b,c) for their study of the photometric 
redshift distribution in the Hubble Deep Field. The reader could also be referred to these works of 
Massarotti \etal for a full comparison of the present theoretical framework with the codes of 
Bruzual and Charlot (1993) and Fioc and Rocca-Volmerange (1997), extensively used in recent extragalactic studies
(see also Buzzoni 1998 for an in-depth discussion in the context of high-redshift observations).
Here I particularly address the relevant UV features of the present theoretical output in the specific context of
this discussion; the complete data set is also available in electronic form at the author's Web 
site.\footnote{{\sf http://www.merate.mi.astro.it/$\sim$eps/home.html}}

A major advantage of our theoretical approach over the previous synthesis codes (with perhaps the
only relevant exception of the Arimoto and Jablonka 1991 models)
resides in the fact that galaxy photometric evolution here comes as a result of the {\it individual} evolution 
of the disk and bulge sub-systems. As a general trend, disk luminosity is expected to slowly increase with 
time (because low-mass stars are ``secularly'' cumulating in a galaxy), as opposed to those in the bulge, which 
will fade due to an increasing fraction of dead stars (Renzini and Buzzoni 1986).
Galaxy  evolution back in time therefore results from the actual balance of these two different photometric trends;
as a consequence, one single SFR characteristic timescale, as usually assumed in other synthesis codes, may not be 
an adequate description of the system as a whole.

\begin{figure}[t]
\resizebox{\hsize}{!}{\includegraphics{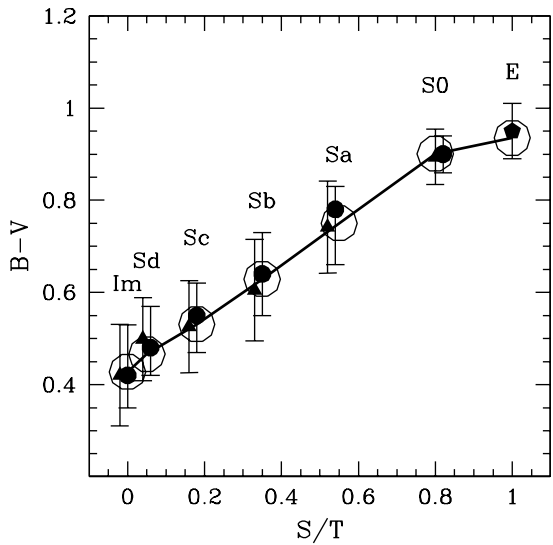}}
\caption{Theoretical colors for template galaxies of different morphological type, according to
Table~\ref{caliball} (big ``$\circ$'' markers), compared with the mean $B-V$ locus from Roberts and 
Haynes (1994) ($\bullet$) and Buta \etal (1994) ($\blacktriangle$). Present-day galaxies are assumed
to be 15 Gyr old. Data for ellipticals are from Buzzoni (1995).}
\label{bvst}
\end{figure}

Luminosity partition between the two basic building blocks of our synthesis models (i.e., disk and spheroid 
components) has been set empirically, relying on the Kent (1985) galaxy decomposition profiles.
These observations have been carried out in the Gunn $r$ band and also provide a confident estimate
of the bolometric partition (Buzzoni 1989). From these data, for each Hubble type I calibrated a bolometric 
morphological parameter defined as $S/T = L{\it (spheroid)}/L{\it (tot)}$.
According to the observations, the luminous mass of the disk remains roughly constant 
along the S0~$\to$~Im sequence, while the bulge luminosity decreases in later type spirals
(Arimoto and Jablonka 1991; Gavazzi 1993).

For the spheroid subsystem, a SSP evolution with solar metallicity has been adopted,
according to Buzzoni (1989, 1995). Disk SFR is in the form of a power law, as discussed in the 
previous section, with stars between 0.1 and 120~M$_\odot$ according to a Salpeter IMF.
Luminosity of the disk sub-system can be computed via eq.~(\ref{eq:lgal}), in which I tune up the SFR index 
$\eta$ so as to reproduce the $B-V$ color of present-day galaxies along the whole Hubble sequence.
The extensive data samples of Roberts and Haynes (1994) and Buta \etal (1994) have been
used as a reference in this regard (cf.\ Fig.~\ref{bvst}).

A slightly subsolar mean metallicity (i.e.\ $<$[Fe/H]$>$ $= -0.5$~dex) is assumed for the disk stellar 
population. This is meant to be a representative average of the whole stellar population at the different epochs.
Our choice agrees with Arimoto and Jablonka (1991), who suggest a luminosity-weighted value of 
$<$[Fe/H]$> \simeq -0.3$ dex for their later-type galaxy models at 15 Gyr. The empirical age-metallicity 
calibration for the Milky Way by Edvardsson \etal (1993) also provides a  value of 
$<$[Fe/H]$> = -0.47$ dex when averaging on current stellar age.
The conclusions of Sec.\ 2.2 assure us, however, that metallicity is not a critical parameter
for model predictions as far as UV evolution is concerned.

\begin{figure}[t]
\resizebox{\hsize}{!}{\includegraphics{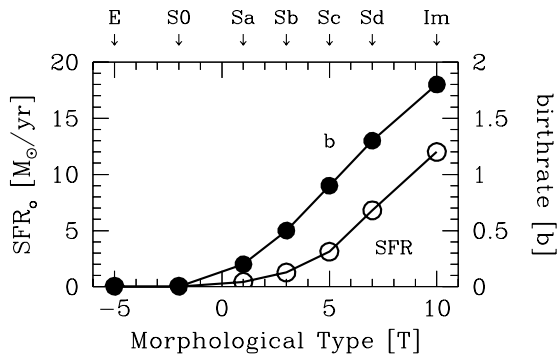}}
\caption{Current SFR for disk stellar populations as expected from our
template models for different morphological types (left scale refers to open dots 
in the plot). A total mass of $10^{11}$~M$_\odot$ is assumed for the galaxies.
Right scale reports the mean birthrate $b = {\it SFR}_o/<{\it SFR}>$ according to 
Table~\ref{caliball} (solid dots).}
\label{birthrate}
\end{figure}

Table~\ref{caliball} gives a global summary of the adopted distinctive parameters for the
different Hubble types at 15 Gyr (cf.\ also Table 1 in Massarotti \etal 2001a for a synopsis with the 
output of other synthesis codes).
Figure~\ref{birthrate} displays the calibration of $b$ vs.\ de Vaucouleurs' ``T'' morphological type,
and the current disk SFR expected for template galaxies of total mass 
M$_{tot} = 10^{11}$~M$_\odot$. These results consistently compare with the
empirical estimate of Kennicutt \etal (1994).

\begin{deluxetable}{lcrccccc}
\tabletypesize{\footnotesize}
\tablecolumns{9}
\tablewidth{0pc}
\tablecaption{Distinctive parameters for 15 Gyr synthesis models}
\tablehead{
\colhead{Hubble Type} & \colhead{$b$\tablenotemark{(a)}} & \colhead{$\eta$} & \colhead{~~$S/T$\tablenotemark{(b)}~~} & \colhead{$M_{disk}/M_{tot}$} & \colhead{$M/L_{bol}$\tablenotemark{(c)}} & \colhead{~~[Fe/H]\tablenotemark{(d)}~~} & \colhead{~~B--V}~~}
\startdata
$\qquad$S0 & 0.0 & $1.0$  & 0.80 & 0.17 & 7.09 & $-0.10$ & 0.90 \\
$\qquad$Sa & 0.2 & $0.8$  & 0.55 & 0.28 & 5.60 & $-0.22$ & 0.75 \\
$\qquad$Sb & 0.5 & $0.5$  & 0.35 & 0.38 & 4.15 & $-0.32$ & 0.63 \\
$\qquad$Sc & 0.9 & $0.1$  & 0.18 & 0.52 & 2.73 & $-0.41$ & 0.53 \\
$\qquad$Sd & 1.3 & $-0.3$  & 0.05 & 0.78 & 1.66 & $-0.48$ & 0.47 \\
$\qquad$Irr & 1.8 & $-0.8$ & 0.00 & 1.00 & 1.11 & $-0.50$ & 0.43 \\
\enddata
\tablenotetext{(a)}{Disk stellar birthrate, $b = {\it SFR}/<{\it SFR}>$}
\tablenotetext{(b)}{$S/T = L{\rm (spheroid)}/L{\rm (tot)}$ in bolometric}
\tablenotetext{(c)}{Bolometric ratio for the whole model galaxy}
\tablenotetext{(d)}{Luminosity-weighted metallicity of the global model from the bolometric partition.}
\label{caliball}
\end{deluxetable}

\begin{figure}[t]
\resizebox{\hsize}{!}{\includegraphics{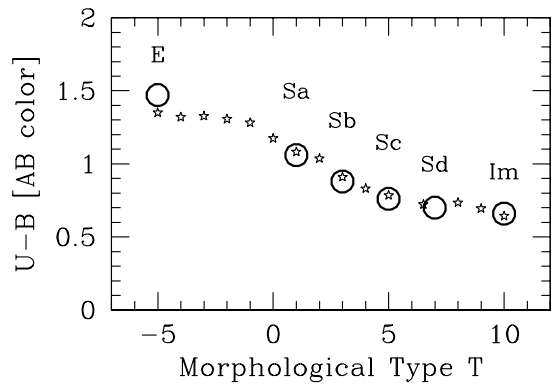}}
\caption{
Mean observed locus for RC3 galaxies of different morphological type
(``$\star$'' markers) according to Buta \etal (1994) compared with
15 Gyr galaxy templates (big ``$\circ$'').
The Johnson $U-B$ color is in AB mag scale. Data have been originally corrected for Galaxy 
and internal average reddening. The typical internal scatter of the observed points is $\pm 0.1$ mag.}
\label{buta}
\end{figure}

\subsection{Model output}

Theoretical evolution of galaxy color and luminosity, pertinent to the different Hubble
morphological types, is summarized in the series of Tables~\ref{e_tab} to \ref{im_tab}.  
All the color entries in the tables are given in AB magnitude scale\footnote{For example, $(U_{20}-B) 
= -2.5\,(\log L_{2000} - \log L_B)$, in the frequency domain, according e.g.\ to Oke and Schild (1970)}
including the Johnson $(U-B)$, while $L_B$ is given in erg~s$^{-1}$Hz$^{-1}$ and $L_{Bol}$ in erg~s$^{-1}$.
Total luminosity of late-type galaxy models is normalized so as to provide a 
disk SFR~$ = 1$~M$_\odot$ yr$^{-1}$ at 15 Gyr; the template elliptical (taken from Buzzoni 1995)
is scaled to a total mass of $5.0\ 10^{11}$~M$_\odot$.
The corresponding value of M$_{tot}$ for the other galaxy templates is reported in each table. 
This value only refers to the amount of mass converted in stars (that is M$_{tot} = \int {\rm SFR}\ dt$) 
and therefore does {\it not} include residual gas. By definition, its value increases with time.
Absolute bolometric magnitude can be derived from the data as ${\it Bol} = -2.5~(\log L_{Bol} -33.59)+4.72$.

\begin{deluxetable}{cccccccc}
\tabletypesize{\footnotesize}
\tablecolumns{8}
\tablewidth{0pc}
\tablecaption{UV-luminosity evolution for E galaxies\tablenotemark{(a)}}
\tablehead{
\colhead{Age} & \colhead{U$_{16}$--B} & \colhead{U$_{20}$--B} & \colhead{U$_{28}$--B} & \colhead{U--B} & \colhead {  } & \colhead{$\log L_B$} & \colhead{$\log L_{Bol}$} \\
\colhead{{\rm [Gyr]}} &    &            &            &     &     &      &    }
\startdata
\phn 1.0& \phn 6.68& 3.76& 2.34& 1.05& & 30.01& 45.29\\
\phn  2.0& \phn 8.90& 5.22& 2.82& 1.16& &  29.72& 45.06\\
\phn  4.0& 10.34& 6.31& 2.89& 1.27& &  29.44& 44.83\\
\phn  5.0& 10.41& 6.48& 2.91& 1.30& &  29.35& 44.76\\
\phn  6.0& 10.25& 6.72& 2.97& 1.33& &  29.27& 44.70\\
\phn  8.0& \phn 9.95& 7.15& 3.08& 1.38& &  29.15& 44.61\\
 10.0& \phn 9.65& 7.57& 3.22& 1.41& &  29.06& 44.53\\
 12.5& \phn 9.12& 7.74& 3.32& 1.45& &  28.97& 44.46\\
 15.0& \phn 8.66& 7.81& 3.41& 1.47& &  28.90& 44.40\\
\enddata
\tablenotetext{(a)}{Colors are in AB scale, including Johnson U--B.
$L_B$ is in erg~s$^{-1}$~Hz$^{-1}$ while $L_{Bol}$ in erg~s$^{-1}$.
The model is normalized to M$_{tot} = 5.0\ 10^{11}$~M$_\odot$.}
\label{e_tab}
\end{deluxetable}

\begin{deluxetable}{cccccccc}
\tabletypesize{\footnotesize}
\tablecolumns{8}
\tablewidth{0pc}
\tablecaption{UV-luminosity evolution for S{\rm a} galaxies\tablenotemark{(a)}}
\tablehead{
\colhead{Age} & \colhead{U$_{16}$--B} & \colhead{U$_{20}$--B} & \colhead{U$_{28}$--B} & \colhead{U--B} & \colhead{  } & \colhead{$\log L_B$} & \colhead{$\log L_{Bol}$} \\
\colhead{{\rm [Gyr]}} &    &            &            &     &     &      &    }
\startdata
\phn 1.0&  2.12&  2.02&  1.87&  0.89& & 29.78& 44.97\\
\phn 2.0&  2.13&  2.13&  2.05&  0.94& & 29.53& 44.77\\
\phn 4.0&  2.13&  2.16&  2.10&  0.98& & 29.29& 44.57\\
\phn 5.0&  2.13&  2.16&  2.11&  1.00& & 29.22& 44.51\\
\phn 6.0&  2.13&  2.17&  2.14&  1.01& & 29.15& 44.45\\
\phn 8.0&  2.14&  2.17&  2.17&  1.03& & 29.05& 44.37\\
 10.0&  2.14&  2.18&  2.21&  1.04& & 28.98& 44.31\\
 12.5&  2.14&  2.18&  2.23&  1.05& & 28.90& 44.24\\
 15.0&  2.15&  2.19&  2.25&  1.06& & 28.84& 44.19\\
\enddata
\tablenotetext{(a)}{
Same entries as in Table~\ref{e_tab} but model total mass assumes\\
SFR = 1 M$_\odot$yr$^{-1}$ or M$_{tot} = 2.26\ 10^{11}$~M$_\odot$ at 15 Gyr.}
\label{sa_tab}
\end{deluxetable}

\begin{deluxetable}{cccccccc}
\tabletypesize{\footnotesize}
\tablecolumns{8}
\tablewidth{0pc}
\tablecaption{UV-luminosity evolution for S{\rm b} galaxies\tablenotemark{(a)}}
\tablehead{
\colhead{Age} & \colhead{U$_{16}$--B} & \colhead{U$_{20}$--B} & \colhead{U$_{28}$--B} & \colhead{U--B} & \colhead {  } & \colhead{$\log L_B$} & \colhead{$\log L_{Bol}$} \\
\colhead{{\rm [Gyr]}} &    &            &            &     &     &     &     }
\startdata
\phn 1.0&  1.73&  1.68&  1.67&  0.84& & 29.27& 44.44\\
\phn 2.0&  1.63&  1.65&  1.73&  0.85& & 29.07& 44.27\\
\phn 4.0&  1.56&  1.60&  1.71&  0.86& & 28.89& 44.11\\
\phn 5.0&  1.54&  1.59&  1.71&  0.87& & 28.84& 44.07\\
\phn 6.0&  1.53&  1.58&  1.71&  0.87& & 28.79& 44.03\\
\phn 8.0&  1.53&  1.57&  1.72&  0.87& & 28.73& 43.97\\
 10.0&  1.52&  1.57&  1.74&  0.87& & 28.68& 43.93\\
 12.5&  1.52&  1.57&  1.74&  0.88& & 28.63& 43.88\\
 15.0&  1.52&  1.57&  1.75&  0.88& & 28.59& 43.85\\
\enddata
\tablenotetext{(a)}{
Same entries as in Table~\ref{e_tab} but model total mass assumes\\
SFR = 1 M$_\odot$yr$^{-1}$  or M$_{tot} = 7.68\ 10^{10}$~M$_\odot$ at 15 Gyr.}
\label{sb_tab}
\end{deluxetable}

\begin{deluxetable}{cccccccc}
\tabletypesize{\footnotesize}
\tablecolumns{8}
\tablewidth{0pc}
\tablecaption{UV-luminosity evolution for S{\rm c} galaxies\tablenotemark{(a)}}
\tablehead{
\colhead{Age} & \colhead{U$_{16}$--B} & \colhead{U$_{20}$--B} & \colhead{U$_{28}$--B} & \colhead{U--B} & \colhead{  } & \colhead{$\log L_B$} & \colhead{$\log L_{Bol}$} \\
\colhead{{\rm [Gyr]}} &    &            &            &     &     &      &    }
\startdata
\phn 1.0&  1.61&  1.58&  1.60&  0.83& & 28.74& 43.92\\
\phn 2.0&  1.36&  1.39&  1.54&  0.80& & 28.61& 43.79\\
\phn 4.0&  1.22&  1.26&  1.45&  0.77& & 28.52& 43.70\\
\phn 5.0&  1.20&  1.24&  1.44&  0.77& & 28.51& 43.68\\
\phn 6.0&  1.19&  1.23&  1.44&  0.76& & 28.49& 43.66\\
\phn 8.0&  1.18&  1.22&  1.44&  0.76& & 28.48& 43.65\\
 10.0&  1.18&  1.23&  1.45&  0.76& & 28.47& 43.64\\
 12.5&  1.19&  1.23&  1.46&  0.76& & 28.46& 43.64\\
 15.0&  1.20&  1.24&  1.47&  0.76& & 28.46& 43.64\\
\enddata
\tablenotetext{(a)}{
Same entries as in Table~\ref{e_tab} but model total mass assumes\\
SFR = 1 M$_\odot$yr$^{-1}$ or M$_{tot} = 3.13\ 10^{10}$~M$_\odot$ at 15 Gyr.}
\label{sc_tab}
\end{deluxetable}

\begin{deluxetable}{cccccccc}
\tabletypesize{\footnotesize}
\tablecolumns{8}
\tablewidth{0pc}
\tablecaption{UV-luminosity evolution for S{\rm d} galaxies\tablenotemark{(a)}}
\tablehead{
\colhead{Age} & \colhead{U$_{16}$--B} & \colhead{U$_{20}$--B} & \colhead{U$_{28}$--B} & \colhead{U--B} & \colhead{  } & \colhead{$\log L_B$} & \colhead{$\log L_{Bol}$} \\
\colhead{{\rm [Gyr]}} &    &            &            &     &     &      &    }
\startdata
\phn 1.0&  1.28&  1.28&  1.41&  0.76& & 28.14& 43.28\\
\phn 2.0&  1.00&  1.05&  1.27&  0.70& & 28.12& 43.23\\
\phn 4.0&  0.93&  0.97&  1.22&  0.67& & 28.18& 43.28\\
\phn 5.0&  0.93&  0.98&  1.22&  0.67& & 28.21& 43.31\\
\phn 6.0&  0.94&  0.98&  1.23&  0.67& & 28.24& 43.34\\
\phn 8.0&  0.96&  1.00&  1.25&  0.68& & 28.28& 43.39\\
 10.0&  0.98&  1.02&  1.27&  0.68& & 28.32& 43.43\\
 12.5&  1.01&  1.05&  1.30&  0.69& & 28.36& 43.48\\
 15.0&  1.03&  1.07&  1.32&  0.70& & 28.39& 43.52\\
\enddata
\tablenotetext{(a)}{
Same entries as in Table~\ref{e_tab} but model total mass assumes\\
SFR = 1 M$_\odot$yr$^{-1}$  or M$_{tot} = 1.45\ 10^{10}$~M$_\odot$ at 15 Gyr.}
\label{sd_tab}
\end{deluxetable}

\begin{deluxetable}{cccccccc}
\tabletypesize{\footnotesize}
\tablecolumns{8}
\tablewidth{0pc}
\tablecaption{UV-luminosity evolution for I{\rm m} galaxies\tablenotemark{(a)}}
\tablehead{
\colhead{Age} & \colhead{U$_{16}$--B} & \colhead{U$_{20}$--B} & \colhead{U$_{28}$--B} & \colhead{U--B} & \colhead{  } & \colhead{$\log L_B$} & \colhead{$\log L_{Bol}$} \\
\colhead{{\rm [Gyr]}} &    &            &            &     &     &       &    }
\startdata
\phn 1.0&  0.42&  0.48&  0.80&  0.50& & 27.21& 42.16\\
\phn 2.0&  0.56&  0.62&  0.91&  0.54& & 27.50& 42.49\\
\phn 4.0&  0.70&  0.75&  1.03&  0.58& & 27.80& 42.82\\
\phn 5.0&  0.74&  0.79&  1.06&  0.60& & 27.89& 42.93\\
\phn 6.0&  0.77&  0.82&  1.09&  0.61& & 27.97& 43.01\\
\phn 8.0&  0.82&  0.87&  1.14&  0.62& & 28.09& 43.15\\
 10.0&  0.86&  0.91&  1.17&  0.64& & 28.18& 43.25\\
 12.5&  0.90&  0.95&  1.21&  0.65& & 28.28& 43.36\\
 15.0&  0.93&  0.98&  1.23&  0.66& & 28.35& 43.44\\
\enddata
\tablenotetext{(a)}{
Same entries as in Table~\ref{e_tab} but model total mass assumes\\
SFR = 1 M$_\odot$yr$^{-1}$ or M$_{tot} = 0.83\ 10^{10}$~M$_\odot$ at 15 Gyr.}
\label{im_tab}
\end{deluxetable}

A preliminary ``sanity check'' of our theoretical output is performed in Fig.~\ref{buta}, which compare with the 
Buta \etal (1994) empirical $U-B$ locus for over 2500 local galaxies in the RC3 catalog (de Vaucouleurs \etal 1991).
It should be pointed out, in this regard, that this check is not completely independent, since I
already used the Buta \etal $B-V$ distribution to constrain the SFR; this would actually
``recover'' any further internal mechanism modulating the galaxy SED.
In any case, it is comforting to see from Fig.~\ref{buta} that the previous $B-V$ vs.\ S/T calibration correctly
accounts, within the internal scatter of the observations, also for the $U-B$ color trend vs.\ ``T'' morphological 
parameter.

\begin{figure}[t]
\resizebox{\hsize}{!}{\includegraphics{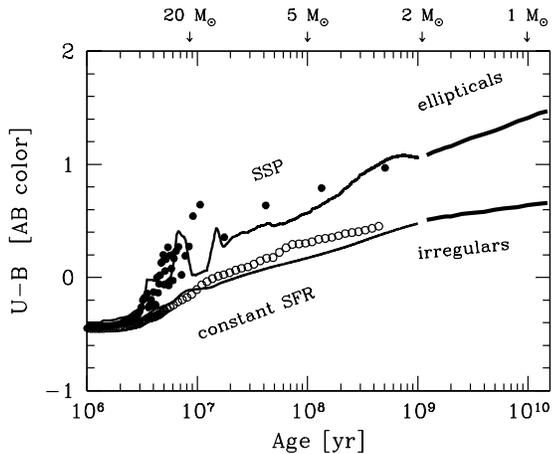}}
\caption{
Theoretical $U-B$ colors (in AB mag scale) for primeval galaxies with continuous star formation 
(``bluer'' models) and SSP evolution (``redder'' models).
Models with $Z_\odot$ and Salpeter IMF from Leitherer 
and Heckman (1995) (open and solid dots), and Leitherer \etal (1999) (thin solid lines) 
are compared with our reference templates for E and Im galaxies 
(thick solid lines, for $t \geq 1$ Gyr). The SSP evolution consistently matches
early-type systems at later epochs while a constant SFR suitably describes late-type and 
irregular galaxies. The SSP TO stellar mass at different epochs is reported on the top scale,
according to eq.~(\ref{eq:fit_clock}).}
\label{lh}
\end{figure}

The models given in Tables~\ref{e_tab}--\ref{im_tab} better focus on ``late'' galaxy evolution, 
after the first Gyr of life, when low-mass stars with M~$\lesssim 2.0$~M$_\odot$ dominate galaxy bolometric 
luminosity, and bulge and disk subsystems begin to differentiate morphology along the Hubble sequence 
(Larson 1975, 1976; Firmani and Tutukov 1992). Evolution is less univocally constrained at earlier epochs, where
convection and mass loss via stellar winds (both depending on metallicity) play an
important role in high-mass stars and affect, among other things, the luminosity partition between 
blue and red giant stars or the relative contribution of Wolf-Rayet stars (see Mass-Hesse and Kunth 1991,
Cervi\~no and Mass-Hesse 1994, and especially Cervi\~no \etal 2002 for a quantitative assessment of these problems 
on the framework of stellar population synthesis).

\begin{figure}[t]
\resizebox{\hsize}{!}{\includegraphics{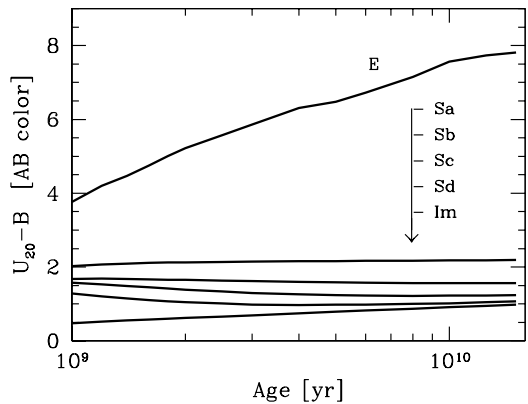}}
\caption{Theoretical rest-frame evolution of the $(U_{20}-B)$ AB color for model galaxies of 
different morphological types.}
\label{evol}
\end{figure}

A complementary in-depth analysis of the first $10^9$ yrs in star-forming galaxies has been carried 
out by Leitherer and Heckman (1995) and Leitherer \etal (1999).
Their models take into account in some detail spectrophotometric evolution of high-mass stars in different 
theoretical environments including continuous star formation and SSP evolution.
A match of our synthesis models and the Leitherer \etal output for $Z_\odot$ and a Salpeter IMF
is shown in Fig.~\ref{lh}. We see that SSP evolution consistently describes early-type systems at primeval 
epochs, while continuous SFR accounts for late-type and irregular galaxies.
It is interesting to note, in addition, that colors of both ellipticals and late-type
galaxies ``degenerate'' in the first few Myr, that is, on a timescale comparable to
$t_{min}$, as expected from eq.~(\ref{eq:lmin}).

Restframe evolution of the $(U_{20}-B)$ AB color for the different galaxy morphological types is displayed in 
Fig.~\ref{evol}. Only ellipticals show a drastic color change becoming sensibly bluer at
earlier epochs; late-type systems remain on the contrary more or less the same, since they are dominated
throughout by fresh star formation.

\begin{figure}[t]
\resizebox{\hsize}{!}{\includegraphics{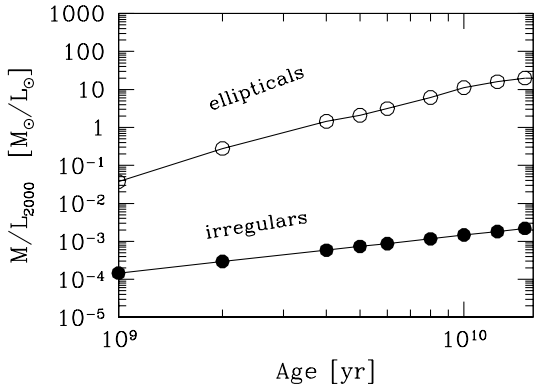}}
\caption{Expected evolution of the galaxy M/L ratio at 2000 \AA\ for the model ellipticals and 
irregulars, according to the data in Table 4 and 9, respectively. Both mass and luminosity are in 
solar units. A solar reference luminosity $L_\odot = 2.39\,10^{15}$~erg~s$^{-1}$~Hz$^{-1}$ 
is assumed at 2000 \AA\ from the Kurucz (1992) model atmospheres.}
\label{ml}
\end{figure}

The theoretical M/L ratio at 2000 \AA\ for model ellipticals and irregulars is explored in Fig.~\ref{ml}.
I have considered the actual total mass of the galaxies, M$_{tot}$, in solar unit according to the previous 
definition, while the 2000 \AA\ luminosity derives from Table 4 and 9 as $\log L_{2000} = \log L_B -0.4\,(U_{20}-B)$.
The value has been converted to solar units assuming $L_\odot = 2.39\,10^{15}$~erg~s$^{-1}$~Hz$^{-1}$
for the Sun at 2000 \AA\ as from a Kurucz (1992) $(T_{\it eff}, \log g) = (5780~{\rm K}, 4.5~{\rm dex})$
model atmosphere.
As a striking feature, note that at 2000 \AA\ enhanced star formation makes Im galaxies nearly 
four orders of magnitude brighter than ellipticals, at comparable total mass.
The role of late-type systems as powerful (and possibly dominant) contributors to the cosmic
UV background has recently been emphasized by Steidel \etal (2001) and Bianchi \etal (2001) through their study of
Lyman-break galaxies.

\section{Comparison with UV observations}

\begin{figure}[t]
\resizebox{\hsize}{!}{\includegraphics{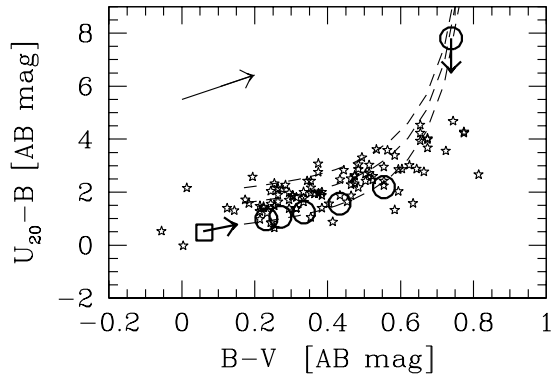}}
\caption{
Two-color diagram of the 149 late-type galaxies in the Donas \etal (1987)
sample (``$\star$'' markers).
Big ``$\circ$'' markers show the locus for 15 Gyr template galaxies
along the sequence Im~$\to$~E, in the sense of increasing $B-V$.
The Im model at 1 Gyr is also displayed (big ``$\sq$'').
Reference models assume a Salpeter IMF with $M_{up} = 120\ M_\odot$.
The effect of residual star formation in the early-type templates is indicated
by the vertical arrow on the E model.
A change in the IMF upper cutoff mass for $M_{up}$ to 80 and $60\ M_\odot$, in the sense 
of ``reddening'' $U_{20}-B$, is explored by the 
dashed curves. All the colors in the plot are in AB mag scale. Galaxy reddening 
is accounted for in the original observations. Reddening vector at the top left is from 
Seaton (1979). The typical error bar for the UV data is $\pm 0.5$ mag on $U_{20}-B$.}
\label{donas1}
\end{figure}

Because of the intrinsic physical limits of atmosphere absorption, a study of 
local galaxies below 3500 \AA\ perforce relies on a quite scanty set of data, mainly 
from IUE observations and balloon-borne missions.
The work of  Donas \etal (1987), collecting 2000 \AA\ photometry for 149 mainly late-type galaxies from
the SCAP 2000 balloon-borne mission, provides an important reference in this regard.
Despite the large uncertainties in the original flux calibration, and the 
somewhat heterogeneous match between $U_{20}$ and (external) $B$ photometry, a comparison with our model
predictions can be attempted in Fig.~\ref{donas1}.
The original data have been corrected for Milky Way extinction, relying on the Burstein and 
Heiles (1984) reddening map to estimate the color excess for each galaxy in the sample.
According to Seaton (1979), the reddening vector is $E(U_{20}-B)/E(B-V) = 4.63$, as displayed in the figure.

Three main features are worth of attention in the plot.
{\it i)} The data distribution spans a wider range in $B-V$, with ``blue'' and ``red'' outliers with respect to the 
15 Gyr late-type model sequence.
{\it ii)} The ``thickness'' of the $U_{20}-B$ data distribution exceeds the internal photometric error 
[reportedly $\sigma(U_{20}-B) = \pm 0.5$ mag], reaching a dispersion of $\pm 0.93$~mag for Sa-Sb types and 
$\pm 0.65$~mag for later types spirals and irregulars, as indicated by Donas \etal (1987).
{\it iii)} The observations tend to be $\sim 0.5$~mag ``redder'' in $U_{20}-B$ than the 
late-type model sequence with M$_{up} = 120$~M$_\odot$.

A careful analysis of the data shows that the three blue outliers 
[i.e.\ those galaxies with $(B-V) \sim 0$ in Fig.~\ref{donas1}] are all Im systems ($T = 9-10$) with active 
star formation. Their integrated colors are consistent with a bulk of early-type stars (i.e.\ of spectral 
type $A5$~V or earlier) and therefore indicate a younger age. This is confirmed by a better fit with our 1 Gyr 
Im model, as reported in the figure. On the other side, the five ``red'' outliers [i.e. those with 
$(B-V) \gtrsim 0.7$ in the plot] are all early-type or bulge-dominated systems, and are in fact ``bluer'' outliers 
when compared with the $U_{20}-B$ color of the early-type template models. These data could easily be accounted for 
by the $E/S0$ templates considering a weak ($\sim 0.2$~M$_\odot$yr$^{-1}$) residual star formation instead of
a plain SSP evolution (as assumed in our calculations, cf.\ Table~\ref{caliball}). Although also marginally 
detectable in the $U-B$ plot (cf.\ Fig.~\ref{buta}), this feature more sensibly affects the 2000 \AA\ luminosity by 
changing the UV vs.\ optical relative emission of the galaxies. In this sense, it is important to stress that UV 
colors of our $E/S0$ models should be taken as redder (upper) limits, as indicated in Fig.~\ref{donas1}.

Concerning point {\it ii)}, Donas \etal (1987) ascribe most of the observed $U_{20}-B$ scatter to
intrinsic galaxy-to-galaxy SFR variations. As expected, the effect should magnify at UV wavelengths,
compared for example to the $B-V$ scatter. According to eq.~(\ref{eq:lmin}), the intrinsic $U_{20}-B$ distribution
implies a SFR dispersion of over a factor of two among galaxies of the same morphological type.
Table~\ref{caliball} helps translating this effect in term of $B-V$ variation. 
By doubling the disk birthrate, for example  between the Sc and the Im template, we have a change of
$\Delta (B-V) = 0.1$ mag in the galaxy integrated color; this leads to a slope of roughly 
$\Delta(U_{20}-B)/\Delta(B-V) \sim 7$ for the star-formation ``color vector'' of late-type systems.

As a final point in our discussion, the ``redder'' offset of the $U_{20}-B$ color distribution with respect to the 
model sequence is open to different explanations (apart from the comparable internal uncertainty in the $U_{20}$ 
photometric zero-point of the observations). According to Table~\ref{zero}, a lower UV emission can be achieved 
either by decreasing the IMF upper cutoff or with a steeper IMF slope (that is assuming a dwarf-dominated 
stellar population); alternatively, we should call for a systematically lower stellar 
birthrate in the galaxies. In any case, the required change both in the IMF slope and in the current SFR would also 
lead to an exceedingly redder $B-V$ (Buzzoni 1989), while a lower value for M$_{up}$ 
could better match the data by selectively reducing the 2000 \AA\ emission alone, as shown in Fig.~\ref{donas1}.

A further and possibly crucial issue in this regard, however, is the galaxy internal extinction;
the presence of dust would in fact act in the sense of a systematic bias toward redder UV colors.
According to the Calzetti (1999) attenuation law, we obtain a color vector $E(U_{20}-B)/E(B-V) = 3.70$, so that
a typical (internal) color excess $E(B-V) \sim 0.1$~mag could fairly account for the observations.

\subsection {Dust and UV absorption}

Our models for galaxy spectral evolution do not explicitly include dust absorption.
Its effect on galaxy SED is in the sense of suppressing UV emission
and enhancing infrared luminosity, as a consequence of the thermalization process.
In addition to stellar extinction, when considering external galaxies we should also 
account for the contribution of grain scattering, as well as the geometry of dust distribution 
and its partition between diffuse and ``clumpy'' phases.  
While scattering could sensibly alleviate the UV extinction (Bruzual \etal 1988; Witt \etal 1992),
a prevailing presence of dust segregated in interstellar clouds would act in 
the opposite sense (Calzetti \etal 1994; Kuchinski \etal 1998). 

A direct way to probe dust attenuation in the galaxy SED is to evaluate the flattening of the
spectral slope at short wavelength, assuming $L(\lambda) \propto \lambda^\beta$ (Kinney \etal 1993).
In agreement with Leitherer's \etal (1999) calculations, our dust-free models predict for late-type 
galaxies a value of $\beta \sim -2.4$ in the wavelength range 1600--2800 \AA\ (cf.\ Tables 5-9).
This should be regarded as a lower limit to the observations, since $\beta$ will in general increase
(up to positive values) in presence of dust absorption.
Note, by the way, that the spectral slope $\beta$ is basically equivalent to a measure of the AB color 
in the relevant wavelength range. From the $U_{16}$ and $U_{28}$ AB magnitudes we have, for instance,
$\beta = 1.65\ (U_{16}-U_{28})-2$.\footnote{By definition, $\beta = \Delta \log F(\lambda)/ \Delta \log \lambda$
and $\log F(\lambda) = \log F(\nu) - 2\,\log (\lambda/c)$. Replacing the AB color definition, 
$C = -2.5\,\Delta \log F(\nu)$, we finally obtain
$\beta = -(0.4/\Delta \log \lambda)\,C -2$, where $\Delta \log \lambda$ is the logarithm 
of the wavelength baseline on which I compute $C$.}

\begin{figure}
\resizebox{\hsize}{!}{\includegraphics{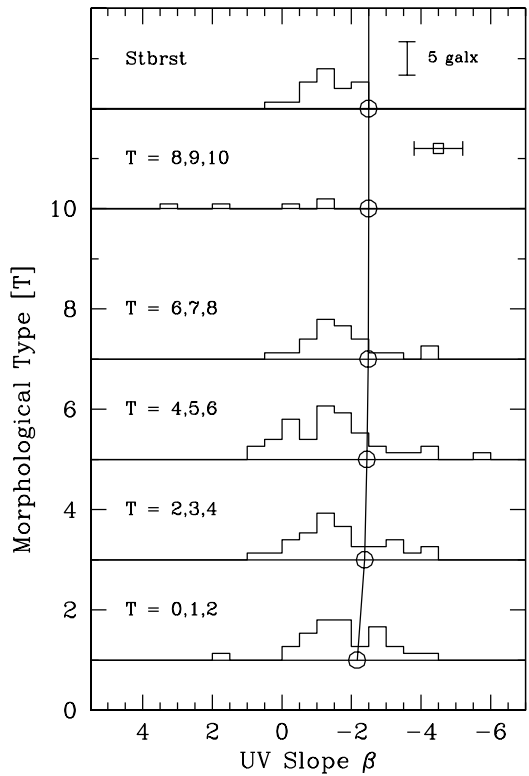}}
\caption{
Observed distribution of the UV spectral slope, $\beta = \Delta \log L (\lambda)/\Delta \log \lambda$,
for 19 starburst galaxies from Gordon \etal (1997) (upper panel) and 93 spirals and irregulars ($T \geq 0$)
from the IUE catalog of Rifatto \etal (1995). The latter sample has been grouped in five bins
according to the morphological parameter $T$, as labeled in the different panels. Only the Gordon \etal data 
have been originally corrected for Galaxy reddening. Open dots and the solid line mark the theoretical lower 
limit to $\beta$ according to our 15 Gyr template models (i.e.\ $\beta = -2.4$).
The reddening vector is $\Delta \beta/ E(B-V) = 3.1$ according to Seaton (1979), while internal attenuation 
is $\Delta \beta/ E(B-V) = 4.5$ following Calzetti (1999). The typical uncertainty in the observed UV slope 
is $\sigma(\beta) \sim \pm 0.7$, for Rifatto \etal (as displayed top right in the plot)
and $\sigma(\beta) \sim \pm 0.25$ for the Gordon \etal galaxies.}
\label{rifatto}
\end{figure}

The spectral slope for local galaxies along the whole Hubble morphological type can be probed relying 
on the catalog of Rifatto \etal (1995). These authors collected homogeneous data for a wide sample of 400 galaxies, 
mainly from the IUE database, reporting magnitudes in three photometric bands at 1650, 2500, and 3150 \AA.
Although not perfectly coincident with our wavelength range, a consistent estimate of $\beta$ 
can be obtained for a total of 161 galaxies from the 1650 and 2500 \AA\ magnitudes. In this regard, we preferred 
not to extrapolate the data to our 1600--2800 \AA\ interval, since only a few galaxies in the 
Rifatto \etal (1995) catalog have confident observations at 3150 \AA, and this would have drastically reduced 
the useful sample for our analysis. Figure~\ref{rifatto} summarizes the observed $\beta$ distribution for the 
late-type galaxy subsample (93 galaxies in total). To study a possible trend with galaxy 
morphology, data have been grouped in five bins according to the de Vaucouleurs' $T$ parameter, as indicated in 
the plots. In addition to the Rifatto \etal (1995) sample, the figure also includes the relevant results for 
19 starburst galaxies from Gordon \etal (1997) with complete 7-band photometry between 
1250 and 2895 \AA; this allowed a confident estimate of $\beta$ at the nominal wavelength 
range of the models.

As a general trend, note from Fig.~\ref{rifatto} that galaxy distribution always peaks, on average,
at flatter UV slopes (i.e.\ $\beta > -2.4$) compared to our theoretical lower limit; this indicates that some 
intervening absorption is systematically affecting the observations, with starburst and later-type galaxies 
(type Sd/Im) marginally more obscured with respect to Sa/Sb systems. The effect of dust reddening seems to
exceed the internal uncertainty of the data, that is, $\sigma (\beta) \simeq \pm 0.7$ for Rifatto \etal (1995) 
and $\sigma (\beta) \simeq \pm 0.25$ for Gordon \etal (1997).
The Gordon \etal (1997) data were originally corrected for the average Galactic extinction, and they 
better track the net effect of internal dust absorption. This is not the case for the Rifatto \etal (1995) sample, 
where the UV flattening is a sum of both internal galaxy attenuation and Milky Way extinction.  

According to the Seaton (1979) extinction law, the expected reddening vector for the 1600--2800 \AA\ UV slope is
$\Delta \beta /E(B-V) = 3.1$, while the Calzetti (1999) attenuation law suggests $\Delta \beta /E(B-V) = 4.5$.
An average flattening of $\Delta \beta \sim 1$, as shown by the data, therefore implies an internal
color excess of $E(B-V) \lesssim 0.2$ mag, consistent with the Donas \etal (1987) data.
This is about a 2 mag extinction at 1600 \AA, according to Calzetti (1999).

\subsection {Dust absorption and high-redshift galaxy evolution}

The role of dust has raised to a central issue in the current cosmological debate, given its importance 
for a fair determination of the cosmic SFR through the UV observation of high-redshift galaxies
(Hughes \etal 1998; Steidel \etal 1999; Massarotti \etal 2001b; Hopkins \etal 2001). 

Generally speaking, a large fraction of residual gas, such as in primeval galaxies, should call 
for a {\it low} amount of dust. The latter is in fact a natural output of stellar evolution 
(mass loss and supernova events actually supply the basic ingredients for grain formation), and
should therefore {\it follow} star formation not {\it precede} it.
As a matter of fact, however, young starbursters certainly appear to be more dust-rich than old 
quiescent ellipticals (Calzetti \etal 1994).
The association of greater quantities of dust with strongly star-forming galaxies may
perhaps be due to a shorter timescale for dust than for star formation, or more likely to a feedback process, 
by which starbursts preferentially occur in dense gas/molecular clouds already heavily contaminated by dust.

Both these arguments point to a tight relationship between dust absorption (and its induced IR emission)
and galaxy UV luminosity, as found by Meurer \etal (1999); apparently, this correlation is already in place
at high redshift, as demonstrated by Adelberger and Steidel (2000).
A dominant contribution of the UV-selected galaxy population to the IR cosmic background is also
in line with the results of Massarotti \etal (2001b), who fully reconcile the SCUBA 
estimates for SFR density at high $z$ (Barger \etal 2001) with the Hubble Deep Field optical observations.

An important point that one should bear in mind, however, is that dust comes from cumulative 
processes inside galaxies, and its content should generally increase with time; contrary to the observations,
this would predict more heavily obscured systems at present time. In order to break this apparent 
dichotomy we need to ``split'', at some point of galaxy evolution, the distribution of the luminous 
matter (i.e.\ stars) and dust. The case of our own galaxy might be explanatory in this sense. 

Because of a higher velocity dispersion and supernova shock waves, young O-B associations leave parent 
molecular clouds and spread their B stars across the 
disk on a timescale of some $10^7$ yrs, a value comparable to the lifetime of stars of
5-10~M$_\odot$ (see, in this regard, the classical work of Becker and Fenkart 1963,
or the recent contribution of Hoogerwerf \etal 2001).
If this is the case also for external galaxies, then an increasing fraction of 
luminous matter would quickly escape the regions of harder and ``clumpy'' dust absorption
({\it \'a la} Calzetti 1999) moving towards a milder and ``scattered'' dust environment 
({\it \'a la} Bruzual \etal 1988). This scenario would consistently explain, among other things, 
why dust attenuation seems to affect more severely gas clouds than stars even in starburst 
galaxies (Calzetti \etal 1994).

\section{Summary and conclusions}

In this paper I have explored some relevant features dealing with the UV luminosity 
evolution of primeval galaxies, in view of a direct application to high-redshift studies.
This has been  done via  a new set of evolutionary population synthesis models that provided
reference templates for different galaxy morphological types.
Theoretical SEDs for late- and early-type systems have been derived, together with
detailed luminosity evolution at 1600, 2000, and 2800 \AA.

As expected, the galaxy UV luminosity below 3000~\AA\ is largely dominated by the
high-mass stellar component, and the distinctive  properties of the integrated SED are
nearly independent of both the total mass of the galaxy and its past star 
formation history. At every age, this leads to a direct relationship between
galaxy UV luminosity and actual SFR (cf.\ eq.~\ref{eq:lmin} and Fig.~\ref{calib}).

Due to a faster evolution of the bulge stellar population (basically a SSP with 
$L_{UV} \propto t^{-\gamma}$  and $\gamma \geq 1$), the high-redshift ancestors of present-day 
spirals should appear as sharply nucleated objects displaying 
scarce or no morphological structure. In addition, early galaxy evolution ($t\leq 1$ Gyr)
could also display some color ``degeneracy'' in the UV range (cf.\ Fig.~\ref{lh}) loosing 
any clear relationship between photometric properties and galaxy morphology along the 
Hubble sequence. This effect should not be neglected for its possible importance in a 
fair sampling of the galaxy population at high redshift (Buzzoni 1998).

The claimed role of late-type systems as prevailing contributors to the cosmic
UV background is reinforced by our results; at 2000 \AA\ Im irregulars are found in fact
nearly four orders of magnitude brighter than ellipticals, per unit luminous mass (cf.\ Fig.~\ref{ml}).

The UV wavelength range is especially sensitive to dust absorption; this  problem
should be properly assessed in high-redshift studies. The SED of local starburst and 
late-type galaxies shows clear evidence of some intrinsic reddening systematically 
affecting galaxies in their active star-forming stage. The amount of luminosity 
extinction for these objects sensibly depends on the assumed dust environment and grain
properties, but it may be as high as 2 mag at 1600 \AA\ in the most typical cases.

Dust is expected to cumulate along the galaxy lifetime, so that the $M_{dust}/L$ ratio 
should increase more than linearly with time for both spirals and ellipticals.
However, the apparent lack of strong reddening in old quiescent systems at present time indicates 
that even in the UV range most of the luminous contribution comes from relatively unabsorbed stars. 
This would call for some evolution in the dust environment, where a pristine ``clumpy'' extinction regime, 
such as in the Calzetti {\it et al.} (1994) scheme, would quickly turn into a diffuse ``scattered'' reddening,
as in the Bruzual {\it et al.} (1988) models.

\acknowledgments

It is a pleasure to thank the anonymous referee for his/her important suggestions that
greatly helped tuning the original discussion. Emanuele Bertone is acknowledged for his precious help
with the calculations of the Kurucz model atmospheres, used in this work.
This project received partial financial support from the Italian MURST under COFIN'98 02-013 and 
COFIN'00 02-016 grants.

\end{document}